\documentclass[12pt]{iopart}
\usepackage{iopams}

\newtheorem{theorem}{Theorem}[section]
\newtheorem{proposition}[theorem]{Proposition}

\newtheorem{corollary}[theorem]{Corollary}

\newlength{\vshift}
\newlength{\hshift}
\usepackage{amssymb,amsopn}

\begin{document}
\begin{flushright}
ICMPA-MPA/027/2012
\end{flushright}
\title{Higher order  SUSY-QM for P\"oschl-Teller potentials: coherent states and operator properties }
\author{Mahouton Norbert Hounkonnou, Sama Arjika and Ezinvi Balo\"itcha}
\address{International Chair of Mathematical Physics
and Applications (ICMPA-UNESCO Chair), University of
Abomey-Calavi, 072 B. P.: 50 Cotonou, Republic of Benin}
\eads{\mailto{norbert.hounkonnou@cipma.uac.bj},
\mailto{rjksama200@gmail.com}
\mailto{ezinvi.baloitcha@cipma.uac.bj}
}

\begin{abstract}
This work prolongs recent investigations by Bergeron et al [see 2012 {\it J. Phys. A: Math. Theo.} {\bf 45} 244028] on 
new SUSYQM coherent states for P\"oschl-Teller potentials.
 It mainly addresses explicit computations of eigenfunctions and spectrum associated to the
 higher order hierarchic supersymmetric  Hamiltonian. Analysis of relevant properties and normal and anti-normal forms is performed and discussed.  Coherent states of the hierarchic first order differential operator $A_{m,\nu,\beta}$ of the P\"oschl-Teller Hamiltonian ${\bf H_{\nu,\beta}^{(m)}}$ and their characteristics 
are studied.
\end{abstract}

 \today

\section{Introduction}
The search for exactly solvable models remains in the core of today research interest in  quantum mechanics.  
A reference list of exactly solvable one-dimensional problems (harmonic oscillator, Coulomb, Morse,
P\"oschl-Teller potentials, etc.)  obtained by an algebraic procedure, namely by a differential operator factorization
methods \cite{infeld}, can be found in \cite{berg} and references therein. This technique, introduced long ago by Schr\"odinger 
\cite{infeld}, was analyzed in depth by Infeld and Hull \cite{ih51}, 
who made an exhaustive classification of factorizable potentials. It was reproduced rather recently in supersymmetric
quantum mechanics (SUSY QM) approach \cite{cooper}
initiated by Witten \cite{wit} 
 and was immediately applied to the hydrogen 
potential \cite{fe84}. This approach gave many new exactly
solvable potentials which were obtained as superpartners of known  exactly solvable models. 
Later 
on, it was  noticed by Witten the possibility of arranging the Schr\"odinger's 
Hamiltonians into isospectral pairs  called {\it supersymmetric partners} \cite{wit}. 
The resulting  supersymmetric quantum mechanics  revived the 
study of exactly solvable Hamiltonians\cite{uh83}.

SUSY QM
is also used for the description
of hidden symmetries of various atomic and nuclear physical systems \cite{gend}.
Besides, it provides a theoretical
laboratory for the investigation of algebraic and dynamical problems in
supersymmetric  field theory.
The simplified setting of SUSY helps to analyze
the difficult problem of dynamical SUSY breaking at full length
and to examine the validity of the Witten index criterion\cite{wit}.

The main result of the present work concerns with the explicit analytical expressions of 
 eigenfunctions and spectrum associated to the
first and
second order  supersymmetric  Hamiltonians with P\"oschl-Teller potentials. The related higher order Hamiltonian coherent states (CS)
 are also constructed and discussed, thus well completing  recent
 investigations   in \cite{berg}  with the same model.

This paper is organized as follows. In Section 2, we recall known results and give an explicit characterization 
of the hierarchic  
Hamiltonians of the  P\"oschl-Teller Hamiltonian ${\bf H_{\nu,\beta}}$. Particular cases of  eigenvalues, eigenfunctions, super-potentials and super-partner potentials are computed.  In Section 3, 
 relevant  operator forms (normal and anti-normal), as well as interesting operator properties and mean-values are discussed.
 In Section 4,  we study the CS related to the first order differential operator $A_{m,\nu,\beta}$ 
of the $m-$ order hierarchic P\"oschl-Teller Hamiltonian ${\bf H_{\nu,\beta}^{(m)}}$ 
and their main mathematical properties, i.e the orthogonality, the normalizability, the continuity in the label and the resolution of the identity. We end with some concluding remarks in Section 5.

\section{The P\"oschl-Teller Hamiltonian and SUSY-QM formalism}
In this section, we first  briefly recall  the 
 P\"oschl-Teller Hamiltonian  model presented in  \cite{berg}. Then, we solve the associated time independent Schr\"odinger equation
with explicit calculation of the wavefunction normalization constant. Finally, from the formalism of higher order hierarchic
supersymmetric factorisation method we derive and discuss main results on the  hierarchy of the  P\"oschl-Teller Hamiltonian.
 
\subsection{The model}
The physical system is described by the Hamiltonian \cite{berg}:
\begin{eqnarray}
\label{ham}
{\bf H_{\nu,\beta}}\phi:=\Big[-\frac{\hbar^2}{2M}\frac{d^2}{dx^2}+V_{\varepsilon_0,\nu,\beta}(x)\Big]
\phi\quad\mbox{for}\quad \phi\in\mathcal{D}_{\bf H_{\nu,\beta}}
\end{eqnarray}
in  a
 suitable  Hilbert space 
 $\mathcal{H}=L^2([0,L],dx)$ endowed with the inner product defined by 
$$(u,v)=\int_{0}^{L}dx\,\bar{u}(x)v(x),\quad u,v\in\mathcal{H},\; [0,L]\subset\mathbb{R}$$
where $\bar{u}$ denotes the complex conjugate of $u.$   $M$ is the particle mass and $\mathcal{D}_{\bf H_{\nu,\beta}}$
 is the domain of definition of ${\bf H_{\nu,\beta}}$. 

\begin{eqnarray}
\label{hamk}
V_{\varepsilon_0,\nu,\beta}(x)=\varepsilon_0
\Big(\frac{\nu(\nu+1)}{\sin^2\frac{\pi x}{L}}-2\beta\cot\frac{\pi x}{L}\Big)
\end{eqnarray}
is the P\"oschl-Teller potential;
 $\varepsilon_0$ is some energy scale, $\nu$ and $\beta$ are  dimensionless parameters. 

 The one-dimensional second-order operator ${\bf H_{\nu,\beta}}$ has singularities at the end points $x=0$ and $x=L$  permiting  to choose $\varepsilon_0\geq 0$
and $\nu\geq 0$. Further, since the symmetry $x\rightarrow L-x$ corresponds to the parameter change $\beta\rightarrow -\beta,$ we can choose $\beta\geq 0.$  As assumed in \cite{berg}, we consider the 
energy scale  
$\varepsilon_0$ as the zero point energy of the energy of the infinite well, i.e. $\varepsilon_0=\hbar^2\pi^2/(2ML^2)$ so that the unique free  parameters of the problem remain
$\nu$ and $\beta$  which will be always assumed to be positive.  The case $\beta=0$ corresponds to the symmetric repulsive potentials investigated in \cite{gazeau}, while the case 
$\beta\neq 0$ leads to the Coloumb potential in the limit $L\rightarrow \infty.$ 

Let us define  the  operator ${\bf \mathcal{H}_{\nu,\beta}}$
with the action $-\frac{\hbar^2}{2M}\phi''(x)+\varepsilon_0
\Big(\frac{\nu(\nu+1)}{\sin^2\frac{\pi x}{L}}-2\beta\cot\frac{\pi x}{L}\Big)\phi$   with the domain being the set of smooth functions with a compact support, $\mathcal{C}_0^\infty(0,L)$.
The P\"oschl-Teller potential is in the limit point case at both ends $x=0$ and $x=L$, if the parameter $\nu\geq 1/2$, and in the limit circle case at both ends if $0\leq \nu< 1/2. $ Therefore, the operator ${\bf \mathcal{H}_{\nu,\beta}}$ is essentially  self-adjoint  in the former case. The closure of ${\bf \mathcal{H}_{\nu,\beta}}$ is 
$\overline{\bf \mathcal{H}_{\nu,\beta}}={\bf H_{\nu,\beta}}$ i.e. $\mathcal{D}_{\overline{\bf \mathcal{H}_{\nu,\beta}}}=\mathcal{D}_{\bf H_{\nu,\beta}}$ and its domain coincides with the maximal one ,
i.e.
$$\mathcal{D}_{{\bf H_{\nu,\beta}}}=\Big\{\phi\in ac^2(0,L),\; \Big[-\frac{\hbar^2}{2M}\phi''+\varepsilon_0
\Big(\frac{\nu(\nu+1)}{\sin^2\frac{\pi x}{L}}-2\beta\cot\frac{\pi x}{L}\Big)\phi\Big]\in\mathcal{H}\Big\},$$ 
 where $ac^2(0,L)$ denotes the absolutely continuous functions with
absolutely continuous derivatives. As mentioned in \cite{berg}, a function of this domain satisfies  Dirichlet boundary conditions 
and in the range of considered $\nu$, 
the deficiency indices of ${\bf \mathcal{H}_{\nu,\beta}}$ is $(2,2)$ indicating that this operator is no longer
essentially self-adjoint but has  a two-parameter family of self-adjoint extensions indeed. As in \cite{berg}, we will restrict only
 to the extension described by Dirichlet boundary conditions, i. e. 
$$\mathcal{D}_{{\bf H_{\nu,\beta}}}=\Big\{\phi\in ac^2(0,L), \mid \phi(0)= \phi(L),\; \Big[-\frac{\hbar^2}{2M}\phi''+\varepsilon_0
\Big(\frac{\nu(\nu+1)}{\sin^2\frac{\pi x}{L}}-2\beta\cot\frac{\pi x}{L}\Big)\phi\Big]\in\mathcal{H}\Big\},$$ 
 $\mathcal{D}_{{\bf H_{\nu,\beta}}}$ is  dense in
$\mathcal{H}$ since $H^{2,2}(0, L)\subset\mathcal{C}_0^\infty(0,L)\subset\mathcal{D}_{{\bf H_{\nu,\beta}}} $ 
and  ${\bf H_{\nu,\beta}}$ is self-adjoint \cite{tesh} where  $H^{m,n}(0, L)$ is the Sobolev space of indice $(m,n)$ \cite{sob}. Later on, we use the dense domain: 
$$\mathcal{D}_{\bf H_{\nu,\beta}}=\Big\{\phi\in AC^2(0,L), \,\varepsilon_0
\Big(\frac{\nu(\nu+1)}{\sin^2\frac{\pi x}{L}}-2\beta\cot\frac{\pi x}{L}\Big)\phi\in\mathcal{H}\Big\},$$
 where $AC_{loc}^2(]0,L[)$ is given by
\begin{eqnarray*}
AC_{loc}^2([0,L])& =&\Big\{\phi\in AC([\alpha,\beta]), \forall \;[\alpha,\beta]\subset \;[0,L],\; [\alpha,\beta] \mbox{ compact }\Big\},\cr
AC[\alpha,\beta]&=&\Big\{\phi\in C[\alpha,\beta], \phi(x)=\phi(\alpha)+\int_{\alpha}^x dt\, g(t), g\in L^1([\alpha,\beta])\Big\}.
\end{eqnarray*}


\subsection{Eigenvalues and eigenfunctions}
The eigen-values $E_{n}^{(\nu,\beta)}$ and functions $\phi_{n}^{(\nu,\beta)}$  solving the  Sturm-Liouville differential 
equation (\ref{ham}),  i.e.
${\bf H_{\nu,\beta}}\phi_{n}^{(\nu,\beta)}=E_{n}^{(\nu,\beta)}\phi_{n}^{(\nu,\beta)},$
 are given by \cite{berg}
\begin{eqnarray}
\label{eigen1}
E_{n}^{(\nu,\beta)}&=&\varepsilon_0\Big((n+\nu+1)^2-\frac{\beta^2}{(n+\nu+1)^2}\Big)\\
\label{eigen2}
\phi_{n}^{(\nu,\beta)}(x)&=&K_n^{(\nu,\beta)}\sin^{\nu+n+1}\frac{\pi x}{L}
\exp \Big(-\frac{\beta \pi x}{L(\nu+n+1)}\Big)P_n^{(a_n,\bar{a}_n)}\Big(i\cot \frac{\pi x}{L}\Big)
\end{eqnarray}
where $n\in\mathbb{N}$, $a_n=-(n+\nu+1)+i\frac{\beta}{n+\nu+1}$, $P_n^{(\lambda,\eta)}(z)$  
are the Jacobi polynomials \cite{ASK} and 
$K_n^{(\nu,\beta)}$ is a normalization constant giving by: 

\begin{eqnarray}
\label{prpos}
K_n^{(\nu,\beta)}
=2^{n+\nu+1}L^{-\frac{1}{2}}
\mathcal{T}(n;\nu,\beta)
\mathcal{O}^{-\frac{1}{2}}(n;\nu,\beta)\exp\Big(\frac{\beta\pi}{2(n+\nu+1)}\Big),
\end{eqnarray}
where
\begin{eqnarray}
\label{rem}
\fl
\mathcal{O}(n;\nu,\beta)&=&\sum_{k=0}^n   \frac{(-n,-2\nu-n-1)_k}{(-\nu-n-\frac{i\beta}{\nu+n+1})_kk!\Gamma(n+\nu+2-k+\frac{i\beta}{\nu+n+1})}\cr
&\times&\sum_{s=0}^n \frac{(-n,-2\nu-n-1,)_s\Gamma(2n+2\nu-s-k+3)}{ (-\nu-n+\frac{i\beta}{\nu+n+1})_s s! \Gamma(n+\nu+2-s-\frac{i\beta}{\nu+n+1})}
\end{eqnarray}
and
\begin{eqnarray}
\label{remo}
\mathcal{T}(n;\nu,\beta)=n!\Big|\Big(-n-\nu+\frac{i\beta}{n+\nu+1}\Big)_n\Big|^{-1}.
\end{eqnarray}
For details on the $K_n^{(\nu,\beta)},$ see Appendix A.

For $n=0$, one can retrieve                                                                                                                                                                                                                                                                                                                                                                                                                                                                                                                                                                                                                                                                                                                                                                                                                                                                                                                                                                                                                                                                                                                                                                                                                             
\begin{eqnarray}
\label{eigen2e}
\phi_0^{(\nu,\beta)}(x)=\frac{2^{\nu+1}}{\sqrt{L\Gamma(2\nu+3)}}
\sin^{\nu+1}\frac{\pi x}{L}
\exp \Big(\frac{\beta \pi }{\nu+1}\Big[\frac{1}{2}-\frac{x}{L}\Big]\Big).
\end{eqnarray}


\subsection{Factorisation method and hierarchy of the  P\"oschl-Teller Hamiltonian: main results }
Let us use the factorization method \cite{cooper,david2, hounk,ih51} to find the hierarchy  of P\"oschl-Teller Hamiltonian. 
We assume the 
ground state eigenfunction $\phi_0^{(\nu,\beta)}$
and eigenvalue $E_0^{(\nu,\beta)}$ are  known. Then we can  
 define the differential operators $A_{\nu,\beta},\; A_{\nu,\beta}^\dag $  factorizing the P\"oschl-Teller  Hamiltonian ${\bf H_{\nu,\beta}}$  (\ref{ham}),  and the associated
 superpotential  $W_{\nu,\beta}$ 
as follows:
\begin{eqnarray}
\label{fact}
{\bf H_{\nu,\beta}}:=\frac{1}{2M}A_{\nu,\beta}^\dag A_{\nu,\beta}+E_0^{(\nu,\beta)},
\end{eqnarray}
where the differential operators $A_{\nu,\beta}$ and $A_{\nu,\beta}^\dag$ are defined by
\begin{eqnarray}
\label{facto}
A_{\nu,\beta}:=\hbar\frac{d}{dx}+W_{\nu,\beta}(x),\quad A_{\nu,\beta}^\dag :=-\hbar\frac{d}{dx}+W_{\nu,\beta}(x),
\end{eqnarray}
acting in  the domains
\begin{eqnarray}
\label{maine}
\fl
\mathcal{D}_{A_{\nu,\beta}}=\{\phi\in ac(0,L)| \;(\hbar \phi'+W_{\nu,\beta}\phi)\in\mathcal{H}\},\\
\fl
\mathcal{D}_{A_{\nu,\beta}^\dag}=\{\phi\in ac(0,L)|\;\exists \;\tilde{\phi}\in \mathcal{H}:
[\hbar\psi(x)\phi(x)]_0^L=0,\;\langle A_{\nu,\beta} \psi,\phi \rangle= \langle
 \psi,\tilde{\phi }\rangle,\;\forall\;\psi\in\mathcal{D}_{A_{\nu,\beta}}\},\nonumber
\end{eqnarray}
where $A_{\nu,\beta}^\dag \phi=\tilde{\phi}$. 
The operator $A_{\nu,\beta}^\dag$ is the adjoint of $A_{\nu,\beta}.$
 Besides,
 considering their common restriction
\begin{eqnarray}
\label{maine}
\mathcal{D}_{A}=\{\phi\in AC(0,L)| \;W_{\nu,\beta}\phi\in\mathcal{H}\},
\end{eqnarray}
we have
$\overline{A_{\nu,\beta}\upharpoonright\mathcal{D}_{A}}=A_{\nu,\beta}$
and $\overline{A_{\nu,\beta}^\dag\upharpoonright\mathcal{D}_{A}}=A_{\nu,\beta}^\dag$. For more details on the role of these operators, see \cite{berg}.
The super-potential $W_{\nu,\beta}$ is 
given by
\begin{eqnarray}
\label{factoz}
 W_{\nu,\beta}(x):=-\hbar\frac{[\phi_0^{(\nu,\beta)}(x)]'}{\phi_0^{(\nu,\beta)}(x)}=-\frac{\pi\hbar}{L}\Big( (\nu+1)\cot\frac{\pi x}{L}-\frac{\beta}{\nu+1}\Big),
\end{eqnarray}
where $\phi_0^{(\nu,\beta)}(x)$ is defined by (\ref{eigen2e}). 
To derive the $m-$th order hierarchic supersymmetric potential, we proceed as follows:

$\bullet$ Permut the operators $A_{\nu,\beta}^\dag$ and 
$A_{\nu,\beta} $ to get  the superpartner Hamiltonian $
{\bf H_{\nu,\beta}^{(1)}}$ of  ${\bf H_{\nu,\beta}:= H_{\nu,\beta}^{(0)}}$:
\begin{eqnarray}
\label{facte}
{\bf H^{(1)}_{\nu,\beta}}:=\frac{1}{2M}A_{\nu,\beta}  A_{\nu,\beta} ^\dag+E_0^{(\nu,\beta)}=-\frac{\hbar^2}{2M}\frac{d^2}{dx^2}+V_{1,\nu,\beta}(x),
\end{eqnarray}
where the partner-potential $V_{1,\nu,\beta}$ of $V_{\nu,\beta}$ is defined by the relation
\begin{eqnarray}
\label{poten}
V_{1,\nu,\beta}(x)&:=&\frac{1}{2M}\Big(W^2_{\nu,\beta}(x)+W'_{\nu,\beta}(x)\Big)+E_0^{(\nu,\beta)}.
\end{eqnarray}
where $W_{\nu,\beta}$ is given by (\ref{factoz}) and $E_0^{(\nu,\beta)}$ by (\ref{eigen1}). In
the equation
\begin{eqnarray}
\label{eqnn}
{\bf H_{\nu,\beta}^{(1)}}\phi_n^{(1,\nu,\beta)}=E_n^{(1,\nu,\beta)}\phi_n^{(1,\nu,\beta)},
\end{eqnarray}
the eigenfunction  $\phi_n^{(1,\nu,\beta)}$ and the eigenvalue $E_n^{(1,\nu,\beta)}$  of ${\bf H_{\nu,\beta}^{(1)}}$ are
related to those of ${\bf H_{\nu,\beta}}$, i.e. $E_n^{(1,\nu,\beta)}:=E_{n+1}^{(\nu,\beta)}$ and $\phi_n^{(1,\nu,\beta)}(x)\propto A_{\nu,\beta}\phi_{n+1}^{(\nu,\beta)}(x)$. 

Since we know $E_0^{(1,\nu,\beta)}$ and $\phi_0^{(1,\nu,\beta)}(x)$,
the Hamiltonian ${\bf H_{\nu,\beta}^{(1)}}$ can be re-factorized to give 
\begin{eqnarray}
\label{eqnrn}
{\bf H_{\nu,\beta}^{(1)}}:=\frac{1}{2M} A_{1,\nu,\beta} ^\dag A_{1,\nu,\beta} +E_0^{(1,\nu,\beta)},
\end{eqnarray}
where
\begin{eqnarray}
\label{factii}
A_{1,\nu,\beta}:=\hbar\frac{d}{dx}+W_{1,\nu,\beta}(x),\quad A_{1,\nu,\beta}^\dag :=-\hbar\frac{d}{dx}+W_{1,\nu,\beta}(x),
\end{eqnarray}
with
\begin{eqnarray}
\label{factii}
W_{1,\nu,\beta}(x):=-\hbar\frac{[\phi_0^{(1,\nu,\beta)}(x)]'}{\phi_0^{(1,\nu,\beta)}(x)}.
\end{eqnarray}

$\bullet$ Permut now the operators $A_{1,\nu,\beta}$ and $A_{1,\nu,\beta}^\dag$ to build 
the third order hierarchic Hamiltonian ${\bf H_{\nu,\beta}^{(2)}},$ i.e. a superpartner Hamiltonian of ${\bf H_{\nu,\beta}^{(1)}}$:
\begin{eqnarray}
\label{ceqnrn}
{\bf H_{\nu,\beta}^{(2)}}:=\frac{1}{2M}  A_{1,\nu,\beta}A_{1,\nu,\beta} ^\dag +E_0^{(1,\nu,\beta)}
=-\frac{\hbar^2}{2M}\frac{d^2}{dx^2}+V_{2,\nu,\beta}(x),
\end{eqnarray}
with
\begin{eqnarray}
V_{2,\nu,\beta}(x)&:=&\frac{1}{2M}\Big(W^2_{1,\nu,\beta}(x)+\hbar W'_{1,\nu,\beta}(x)\Big)+E_0^{(1,\nu,\beta)},
\end{eqnarray}
where $W_{1,\nu,\beta}$ is defined in (\ref{factii}) and $E_0^{(1,\nu,\beta)}$ in (\ref{eigen1}).
Start now from the following equation
\begin{eqnarray*}
{\bf H_{\nu,\beta}^{(2)}}\phi_n^{(2,\nu,\beta)}=E_n^{(2,\nu,\beta)}\phi_n^{(2,\nu,\beta)}.
\end{eqnarray*}
The eigenvalue $E_n^{(2,\nu,\beta)}$ and the eigenfunction    $\phi_n^{(2,\nu,\beta)}$  of $\;{\bf H_{\nu,\beta}^{(2)}}$ are related to those of ${\bf H_{\nu,\beta}^{(1)}}$, i.e. $E_n^{(2,\nu,\beta)}:=E_{n+1}^{(1,\nu,\beta)}$ and $\phi_n^{(2,\nu,\beta)}(x)\propto A_{1,\nu,\beta}\phi_{n+1}^{(1,\nu,\beta)}(x)$.

From known $E_0^{(2,\nu,\beta)}$ and $\phi_0^{(2,\nu,\beta)}$, we can 
 re-factorize the Hamiltonian ${\bf H_{\nu,\beta}^{(2)}}$:
\begin{eqnarray}
\label{cqnrn}
{\bf H_{\nu,\beta}^{(2)}}:=\frac{1}{2M}  A_{2,\nu,\beta}^\dag A_{2,\nu,\beta}  +E_0^{(2,\nu,\beta)}
=-\frac{\hbar^2}{2M}\frac{d^2}{dx^2}+V_{2,\nu,\beta}(x),
\end{eqnarray}
where the operators $A_{2,\nu,\beta}$ and $A_{2,\nu,\beta}^\dag$ are given, respectively, by
\begin{eqnarray}
A_{2,\nu,\beta}:=\hbar\frac{d}{dx}+W_{2,\nu,\beta}(x),\, A_{2,\nu,\beta}^\dag :=-\hbar\frac{d}{dx}+W_{2,\nu,\beta}(x),
\end{eqnarray}
with the superpotential
\begin{eqnarray}
\label{are}
 W_{2,\nu,\beta}(x)=-\hbar\frac{[\phi_0^{(2,\nu,\beta)}(x)]'}{\phi_0^{(2,\nu,\beta)}(x)},
\end{eqnarray}
and the partner potential $V_{2,\nu,\beta}$
\begin{eqnarray}
V_{2,\nu,\beta}(x)&:=&\frac{1}{2M}\Big(W^2_{2,\nu,\beta}(x)-\hbar W'_{2,\nu,\beta}(x)\Big)+E_0^{(2,\nu,\beta)},
\end{eqnarray}
where $W_{2,\nu,\beta}$ is defined in (\ref{are}) and $E_0^{(2,\nu,\beta)}$ in (\ref{eigen1}).

$\bullet$ So, we have shown that one can determine the superpartner Hamiltonian ${\bf H_{\nu,\beta}^{(1)}}$ of ${\bf H_{\nu,\beta}},$ re-factorize ${\bf H_{\nu,\beta}^{(1)}}$
in order to determine its superpartner ${\bf H_{\nu,\beta}^{(2)}},$ then re-factorize ${\bf H_{\nu,\beta}^{(2)}}$  to determine its superpartner ${\bf H_{\nu,\beta}^{(3)}},$
and so on. Each Hamiltonian has  eigenfunctions   and  eigenvalues.
Thus, if the first Hamiltonian $
{\bf H_{\nu,\beta}}$
has $r$ eigenfunctions $\phi_n^{(\nu,\beta)}$  related to the eigenvalues $E_n^{(\nu,\beta)},$ 
$0\leq  n \leq (r-1),$ then one can always generate an hierarchy of $(r-1)$ Hamiltonians ${\bf H_{\nu,\beta}^{(2)}}, {\bf H_{\nu,\beta}^{(3)}},\ldots, {\bf H_{\nu,\beta}^{(r)}}$
such that ${\bf H_{\nu,\beta}^{(m)}}$ has the same eigenvalues as ${\bf H_{\nu,\beta}}$, except for the first $(m-1)$ eigenvalues of ${\bf H_{\nu,\beta}}.$
 In fact, for $m=2,3,4,\ldots,r$, we define  the Hamiltonian in its factorized form as follows:
\begin{eqnarray}
\label{wcqnrn}
{\bf H_{\nu,\beta}^{(m)}}:=\frac{1}{2M}  A_{m,\nu,\beta}^\dag A_{m,\nu,\beta}  +E_0^{(m,\nu,\beta)}
=-\frac{\hbar^2}{2M}\frac{d^2}{dx^2}+V_{m,\nu,\beta}(x),
\end{eqnarray}
while its super-partner Hamiltonian ${\bf H_{\nu,\beta}^{(m+1)}}$ is given by
\begin{eqnarray}
\label{superp}
{\bf H_{\nu,\beta}^{(m+1)}}:&=&\frac{1}{2M} A_{m,\nu,\beta} A_{m,\nu,\beta}^\dag   +E_0^{(m,\nu,\beta)}=-\frac{\hbar^2}{2M}\frac{d^2}{dx^2}+V_{m+1,\nu,\beta}(x),
\end{eqnarray}
 where the operators $A_{m,\nu,\beta}$ and $A_{m,\nu,\beta}^\dag$ are defined by
\begin{eqnarray}
\label{toutt}
A_{m,\nu,\beta}:=\hbar\frac{d}{dx}+W_{m,\nu,\beta}(x),\quad A_{m,\nu,\beta}^\dag: =-\hbar\frac{d}{dx}+W_{m,\nu,\beta}(x),
\end{eqnarray}
which do not commute with the Hamiltonians ${\bf H_{\nu,\beta}^{(m)}}$ and ${\bf H_{\nu,\beta}^{(m+1)}},$ but  satisfy the intertwining relations
\begin{eqnarray}
\label{comm}
{\bf H_{\nu,\beta}^{(m)}}A_{m,\nu,\beta}^\dag=A_{m,\nu,\beta}^\dag{\bf H_{\nu,\beta}^{(m+1)}},\quad 
{\bf H_{\nu,\beta}^{(m+1)}}A_{m,\nu,\beta}=A_{m,\nu,\beta}{\bf H_{\nu,\beta}^{(m)}}.
\end{eqnarray}
The super-potential $ W_{m,\nu,\beta}$ is given by definition by the relation:
\begin{eqnarray}
 W_{m,\nu,\beta}(x):=-\hbar\frac{[\phi_0^{(m,\nu,\beta)}(x)]'}{\phi_0^{(m,\nu,\beta)}(x)},
\end{eqnarray}
while the potential $V_{m,\nu,\beta}$ and its superpartner potential $V_{m+1,\nu,\beta}$ are defined by
\begin{eqnarray}
V_{m,\nu,\beta}(x)&:=&\frac{1}{2M}\Big(W^2_{m,\nu,\beta}(x)-\hbar W'_{m,\nu,\beta}(x)\Big)+E_0^{(m,\nu,\beta)},\cr
V_{m+1,\nu,\beta}(x)&:=&\frac{1}{2M}\Big(W^2_{m,\nu,\beta}(x)+\hbar W'_{m,\nu,\beta}(x)\Big)+E_0^{(m,\nu,\beta)}.
\end{eqnarray}
The energy spectrum $E_n^{(m+1,\nu,\beta)}$ and the  eigenfunction $\phi_{n}^{(m+1,\nu,\beta)}$ of 
the super-partner Hamiltonian ${\bf H_{\nu,\beta}^{(m+1)}}$ are  related  to  those of ${\bf H_{\nu,\beta}^{(m)}}$, i. e. $E_n^{(m+1,\nu,\beta)}:=E_{n+1}^{(m,\nu,\beta)}$ and $\phi_{n}^{(m+1,\nu,\beta)}(x)\propto A_{m,\nu,\beta}\phi_{n+1}^{(m,\nu,\beta)}(x)$ as formulated below.
\begin{proposition}
\label{ppp}
The eigen-energy spectrum $E_n^{(m+1,\nu,\beta)}$  and eigen-function $\phi_{n}^{(m+1,\nu,\beta)}$ that solve the time-independent Schr\"odinger equation for the $(m+1)-$order  hierarchic superpartner Hamiltonian 
$\;{\bf H_{\nu,\beta}^{(m+1)}}$, i.e $\;{\bf H_{\nu,\beta}^{(m+1)}}\phi_{n}^{(m+1,\nu,\beta)}=E_n^{(m+1,\nu,\beta)}\phi_{n}^{(m+1,\nu,\beta)},$ are given, respectively, by: 
\begin{eqnarray}
E_n^{(m+1,\nu,\beta)}&=&\varepsilon_0\Big((n+m+\nu+2)^2-\frac{\beta^2}{(n+m+\nu+2)^2}\Big),
\end{eqnarray}
\begin{eqnarray}
\label{tote}
\phi_{n}^{(m+1,\nu,\beta)}(x)=\frac{A_{m,\nu,\beta}A_{m-1,\nu,\beta}\ldots A_{1,\nu,\beta}A_{\nu,\beta}\phi_{n+m+1}^{(\nu,\beta)}(x)}{\sqrt{(2M)^{m+1}\prod_{k=0}^{m}\Big(E_{n+m+1}^{( \nu,\beta)}-E_k^{( \nu,\beta)}\Big)}}.
\end{eqnarray}
\end{proposition}
As a matter of explicit computation, for the particular value of $m=0,$ we get
\begin{enumerate}
\item the energy spectrum 
\begin{eqnarray}
E_n^{(1,\nu,\beta)}&=&\varepsilon_0\Big((n+\nu+2)^2-\frac{\beta^2}{(n+\nu+2)^2}\Big),
\end{eqnarray}
\item the eigenfunction
\begin{eqnarray}
\fl
\phi_{n}^{(1,\nu,\beta)}(x)=\frac{2^{n+\nu+2}e^{\frac{\beta\pi}{2(n+\nu+2)}}
\mathcal{T}(n+1;\nu,\beta)
}{\sqrt{2ML(n+1)\mathcal{O}(n+1;\nu,\beta)\Delta_{n+1}^0 E_0^{(\nu,\beta)}}}\Bigg[\Bigg[\frac{2M(n+1)^2\overline{\Delta_{n+1}^0 E_0^{(\nu,\beta)}}}{n+2\nu+3}\Bigg]^{1/2}\cr
\times\cos \Big(\frac{\pi x}{L}
-\alpha_{\nu,\beta}(n)\Big)P_{n+1}^{(a_{n+1},\bar{a}_{n+1})}\Big(i\cot \frac{\pi x}{L}\Big)+\frac{i\pi\hbar(n+2\nu+2)}{2L\sin\frac{\pi x}{L}}\cr
\times P_{n}^{(a_{n+1}+1,\bar{a}_{n+1}+1)}\Big(i\cot \frac{\pi x}{L}\Big)\Bigg]
\sin^{\nu+n+1}\frac{\pi x}{L}\exp \Big(-\frac{\beta \pi x}{L(\nu+n+2)}\Big).
\end{eqnarray}
\item the superpotential
\begin{eqnarray}
W_{1,\nu,\beta}(x)&=&-\frac{\hbar\pi}{L}
\Big((\nu+2)\cot\frac{\pi x}{L}-\frac{\beta}{\nu+2}\Big),
\end{eqnarray}
\item the potential
\begin{eqnarray}
V_{1,\nu,\beta}(x)&=&\varepsilon_0
\Big(\frac{(\nu+1)(\nu+2)}{\sin^2\frac{\pi x}{L}}-2\beta\cot\frac{\pi x}{L}\Big),
\end{eqnarray}
where
\begin{eqnarray}
\fl 
\alpha_{\nu,\beta}(n)=\arctan\Bigg(\frac{\beta}{(\nu+1)(\nu+n+2)}\Bigg),\qquad \overline{\Delta_{n+1}^0 E_0^{(\nu,\beta)}}:=\frac{E_{n+1}^{(\nu,\beta)}-E_0^{(\nu,\beta)}}{n+1}.
\end{eqnarray}
\end {enumerate}
 
It is worth noticing that the
 potentials $V_{\varepsilon_0,\nu,\beta}$ and  
$V_{m+1,\nu,\beta}$ are related in a simpler way, i.e  
\begin{eqnarray}
V_{m+1, \nu,\beta}(x)=V_{\varepsilon_0, \nu,\beta}(x)-
\frac{\hbar^2(m+1)(2\nu+m+2)}{2M}
\frac{d^2}{dx^2  }\ln\Big(\sin\frac{\pi x}{L}\Big).
\end{eqnarray}


\section{Relevant operator properties}
This section is devoted to the investigation of relevant properties of the operators 
 ${\bf H_{\nu,\beta}^{(m+1)}}$
 and ${\bf H_{\nu,\beta}}.$
\begin{proposition}
\label{propae3}
For the operators $A_{m,\nu,\beta}$ and $A_{m,\nu,\beta}^\dag$, there is a pair
of $(m+1)-$ order hierarchic operators intertwining
 ${\bf H_{\nu,\beta}}$ and ${\bf H_{\nu,\beta}^{(m+1)}}$, namely
\begin{eqnarray}
\label{46}
{\bf H_{\nu,\beta}}B_m^\dag =B_m^\dag {\bf H_{\nu,\beta}^{(m+1)}},\quad B_m{\bf H_{\nu,\beta}} ={\bf H_{\nu,\beta}^{(m+1)}}B_m,
\end{eqnarray}
where 
\begin{eqnarray}
\label{seu}
B_m:=A_{m,\nu,\beta}\ldots A_{1,\nu,\beta}A_{\nu,\beta},\;\; B_m^\dag:=A_{\nu,\beta}^\dag A_{1,\nu,\beta}^\dag\ldots A_{m,\nu,\beta}^\dag. 
\end{eqnarray}
\end{proposition}
{\bf Proof.} \;By multiplying on the left hand the  intertwining relation (\ref{comm})  by the operator $A_{m-1,\nu,\beta}^\dag$, we have 
\begin{eqnarray*}
A_{m-1,\nu,\beta}^\dag{\bf H_{\nu,\beta}^{(m)}}A_{m,\nu,\beta}^\dag =A_{m-1,\nu,\beta}^\dag A_{m,\nu,\beta}^\dag {\bf H_{\nu,\beta}^{(m+1)}}
\end{eqnarray*}
which is equivalent to
\begin{eqnarray}
\label{proor}
{\bf H_{\nu,\beta}^{(m-1)}}A_{m-1,\nu,\beta}^\dag A_{m,\nu,\beta}^\dag =A_{m-1,\nu,\beta}^\dag A_{m,\nu,\beta}^\dag {\bf H_{\nu,\beta}^{(m+1)}}.
\end{eqnarray}
By continuing the process until the order $m-1$, we have
\begin{eqnarray}
\label{proorf}
{\bf H_{\nu,\beta}^{(1)}}A_{1,\nu,\beta}^\dag\ldots  A_{m,\nu,\beta}^\dag = A_{1,\nu,\beta}^\dag\ldots  A_{m,\nu,\beta}^\dag {\bf H_{\nu,\beta}^{(m+1)}}.
\end{eqnarray}
By multiplying on the left hand the equation (\ref{proorf})  by the operator $A_{\nu,\beta}^\dag$ we have
\begin{eqnarray*}
\label{proorfe}
A_{\nu,\beta}^\dag{\bf H_{\nu,\beta}^{(1)}}A_{1,\nu,\beta}^\dag\ldots  A_{m,\nu,\beta}^\dag = A_{\nu,\beta}^\dag A_{1,\nu,\beta}^\dag\ldots  A_{m,\nu,\beta}^\dag {\bf H_{\nu,\beta}^{(m+1)}}
\end{eqnarray*}
which is equivalent to ${\bf H_{\nu,\beta}}B_m^\dag =  B_m^\dag {\bf H_{\nu,\beta}^{(m+1)}}$.
Similarly we get $B_m {\bf H_{\nu,\beta}} =   {\bf H_{\nu,\beta}^{(m+1)}}B_m$.$\square$

\begin{proposition}
\label{ttttt}
For any positive integers $n, m,$  the following result holds:
\begin{eqnarray}
\label{5dp1ed}
B_nB_m^\dag&=&\left\{\begin{array}{ll}(2M)^{m+1}{_{m+1}\Lambda}_{n,\nu,\beta}\prod_{k=0}^{m}\Big({\bf H_{\nu,\beta}^{(m+1)}}-E_k^{(\nu,\beta)}\Big) &n > m\\\\
(2M)^{n+1}\prod_{k=0}^{n}\Big({\bf H_{\nu,\beta}^{(n+1)}}-E_k^{(\nu,\beta)}\Big)
{^{n+1}\Theta}_{m,\nu,\beta} &n < m. \end{array}
\right.
\end{eqnarray}
In particular, if $n = m$, we have
\begin{eqnarray}
\label{5dp}
B_mB_m^\dag&=&(2M)^{m+1}\prod_{k=0}^{m}\Big({\bf H_{\nu,\beta}^{(m+1)}}-E_k^{(\nu,\beta)}\Big),\\
\label{5dpp}
B_m^\dag B_m&=&(2M)^{m+1}\prod_{k=0}^{m}\Big({\bf H_{\nu,\beta}}-E_k^{(\nu,\beta)}\Big),
\end{eqnarray}
where the operators $ {_{m+1}\Lambda}_{n,\nu,\beta}$ and ${^{n+1}\Theta}_{m,\nu,\beta}$  are given by
\begin{eqnarray}
\label{5dedp}
\fl
{_{m+1}\Lambda}_{n,\nu,\beta}:=A_{n,\nu,\beta}A_{n-1,\nu,\beta}\ldots A_{m+1,\nu,\beta},\;\;
{^{n+1}\Theta}_{m,\nu,\beta}:=A_{n+1,\nu,\beta}^\dag A_{n+2,\nu,\beta}^\dag\ldots A_{m,\nu,\beta}^\dag.
\end{eqnarray}
\end{proposition}

{\bf Proof.}  From (\ref{seu}), we have
\begin{eqnarray*}
B_nB_m^\dag&=&A_{n,\nu,\beta} A_{n-1,\nu,\beta}\ldots  (A_{\nu,\beta}A_{\nu,\beta}^\dag) A_{1,\nu,\beta}^\dag\ldots  A_{m,\nu,\beta}^\dag\cr
&=&2MA_{m,\nu,\beta} \ldots A_{1,\nu,\beta}({\bf H_{\nu,\beta}^{(1)}}-E_0^{(\nu,\beta)})A_{1,\nu,\beta}^\dag\ldots  A_{m,\nu,\beta}^\dag\cr
&=&2MA_{n,\nu,\beta} \ldots A_{1,\nu,\beta}A_{1,\nu,\beta}^\dag({\bf H_{\nu,\beta}^{(2)}}-E_0^{(\nu,\beta)})A_{2,\nu,\beta}^\dag\ldots  A_{m,\nu,\beta}^\dag\cr
&\vdots&\cr
&=&(2M)^{m+1}{_{m+1}\Lambda}_{n,\nu,\beta}\prod_{k=0}^{m}\Big({\bf H_{\nu,\beta}^{(m+1)}}-E_k^{(\nu,\beta)}\Big)
\end{eqnarray*}
if $n>m$,
\begin{eqnarray*}
B_nB_m^\dag&=&A_{n,\nu,\beta} A_{n-1,\nu,\beta}\ldots  (A_{\nu,\beta}A_{\nu,\beta}^\dag) A_{1,\nu,\beta}^\dag\ldots  A_{m,\nu,\beta}^\dag\cr
&=&2MA_{m,\nu,\beta} \ldots A_{1,\nu,\beta}({\bf H_{\nu,\beta}^{(1)}}-E_0^{(\nu,\beta)})A_{1,\nu,\beta}^\dag\ldots  A_{m,\nu,\beta}^\dag\cr
&=&2MA_{n,\nu,\beta} \ldots A_{1,\nu,\beta}A_{1,\nu,\beta}^\dag({\bf H_{\nu,\beta}^{(2)}}-E_0^{(\nu,\beta)})A_{2,\nu,\beta}^\dag\ldots  A_{m,\nu,\beta}^\dag\cr
&\vdots&\cr
&=&(2M)^{n+1}\prod_{k=0}^{n}\Big({\bf H_{\nu,\beta}^{(n+1)}}-E_k^{(\nu,\beta)}\Big)
{^{n+1}\Theta}_{m,\nu,\beta}
\end{eqnarray*}
if $n<m$. \\
For $n=m$, the proof is immediate. $\square$

\begin{corollary}
\label{colo}
The operators ${_{m+1}\Lambda}_{n,\nu,\beta}$ and ${^{n+1}\Theta}_{m,\nu,\beta}$ satisfies the following identities
\begin{eqnarray}
\label{4etttp7}
{_{m+1}\Lambda}_{n,\nu,\beta}\,{_{m+1}\Lambda}_{n,\nu,\beta}^\dag&=&
(2m)^{n-m}\prod_{k=m+1}^{n}\Big({\bf H_{\nu,\beta}^{(n+1)}}-E_{k}^{(\nu,\beta)}\Big),\\
{_{m+1}\Lambda}_{n,\nu,\beta}^\dag\,{_{m+1}\Lambda}_{n,\nu,\beta}&=&
(2m)^{n-m}\prod_{k=m+1}^{n}\Big({\bf H_{\nu,\beta}^{(m+1)}}-E_{k}^{(\nu,\beta)}\Big),\\
\label{4ettt7z}
 {^{n+1}\Theta}_{m,\nu,\beta} {^{n+1}\Theta}_{m,\nu,\beta}^\dag&=&
(2m)^{m-n}\prod_{k=n+1}^{m}\Big({\bf H_{\nu,\beta}^{(n+1)}}-E_{k}^{(\nu,\beta)}\Big),\\
 {^{n+1}\Theta}_{m,\nu,\beta}^\dag {^{n+1}\Theta}_{m,\nu,\beta}&=&
(2m)^{m-n}\prod_{k=n+1}^{m}\Big({\bf H_{\nu,\beta}^{(m+1)}}-E_{k}^{(\nu,\beta)}\Big).
\end{eqnarray}
\end{corollary}
{\bf Proof.} The proof is obviously true by using (\ref{comm}) and (\ref{5dedp}). $\square$

Besides, considering the supercharges
$$ Q:=
\left(\begin{array}{cc}0 & 0\\ B_m& 0\end{array}\right),\quad  Q^\dag:=\left(\begin{array}{cc}0 & B_m^\dag\\
0 & 0\end{array}\right). $$
and  the SUSY Hamiltonian ${\bf H_{\nu,\beta}^{ss}}$  given by
\begin{eqnarray}
\label{susyh}
\fl
{\bf H_{\nu,\beta}^{ss}}:=
(2M)^{m+1}\Bigg[\begin{array}{cc}\prod_{k=0}^{m}({\bf H_{\nu,\beta}}-E_k^{(\nu,\beta)}) & 0\\
0 &\prod_{k=0}^{m}\Big({\bf H_{\nu,\beta}}-E_k^{(\nu,\beta)}+\frac{\varepsilon_0(m+1)(2\nu+m+2)}{\sin^2\frac{\pi x}{L}}\Big)\end{array}\Bigg],
\end{eqnarray}
we readily check, like in \cite{david2, hounk,  wit}, that

\begin{eqnarray}
\label{algera}
{\bf H_{\nu,\beta}^{ss}}=\left\{Q,Q^\dag\right\}:=QQ^\dag+Q^\dag Q,\quad [{\bf H_{\nu,\beta}^{ss}},Q]:={\bf H_{\nu,\beta}^{ss}}Q-Q{\bf H_{\nu,\beta}^{ss}}=0\cr
 [{\bf H_{\nu,\beta}^{ss}},Q^\dag]:={\bf H_{\nu,\beta}^{ss}}Q^\dag-Q^\dag{\bf H_{\nu,\beta}^{ss}}=0.
\end{eqnarray}
In terms of the Hermitian supercharges $$Q_1:=\frac{1}{2}\left(\begin{array}{cc}0 & B_m^\dag \\
B_m&0 \end{array}\right) \quad\; and \;\quad  Q_2:=\frac{1}{2i}\left(\begin{array}{cc}0 & B_m^\dag \\
-B_m&0 \end{array}\right)  $$
 the 
superalgebra (\ref{algera}) takes the form
\begin{eqnarray}
\label{49}
[Q_i, {\bf H_{\nu,\beta}^{ss}}]=0,\quad \{Q_i, Q_j\}:=Q_iQ_j+Q_iQ_i=
\delta_{ij}{\bf H_{\nu,\beta}^{ss}}, \quad i,j=1,2.
\end{eqnarray}
\begin{proposition}
\label{prop4o}
The actions of  $B_m^\dag$ and  $B_m$ on the normalized eigenfunctions $\phi_n^{(m+1,\nu,\beta)}$ and  $\phi_n^{(\nu,\beta)}$ of $\;{\bf H_{\nu,\beta}^{(m+1)}}$ and $\;{\bf H_{\nu,\beta}},$
associated to the eigenvalues $E_{n}^{(m+1,\nu,\beta)}$ and $E_n^{(\nu,\beta)},$
are given by
\begin{eqnarray}
\label{48}
\fl 
B_m^\dag  \phi_n^{(m+1,\nu,\beta)}(x)
&=2^{n+m+\nu+2}(\hbar\pi L^{-1})^{m+1}\mathcal{T}(n+m+1;\nu,\beta)\Big[L\mathcal{O}(n+m+1;\nu,\beta)\Big]^{-\frac{1}{2}}\cr
&\times
\exp\Big[\frac{\beta\pi}{n+m+\nu+2}\Big(\frac{1}{2}-\frac{x}{L}\Big)\Big]\mathcal{M}(n,m;\nu,\beta)\sin^{n+m+\nu+2}\Big(\frac{\pi x}{L}\Big)\cr
&\times P_{n+m+1}^{(a_{n+m+1},\bar{a}_{n+m+1})}\Big(i\cot \frac{\pi x}{L}\Big),
\end{eqnarray}
and
\begin{eqnarray}
\label{cr}
B_m\phi_{n+m+1}^{(\nu,\beta)}(x)= (\pi\hbar L^{-1})^{m+1}\mathcal{M}(n,m;\nu,\beta)
\phi_n^{(m+1,\nu,\beta)}(x),
\end{eqnarray}
respectively, 
where $\mathcal{M}(n,m;\nu,\beta)$ is expressed by
\begin{eqnarray}
\fl
\mathcal{M}^2(n,m;\nu,\beta)
=\prod_{k=0}^{m}\frac{(n+m-k+1)}{(n+m+2\nu+k+3)^{-1}}
\Big(1+\frac{\beta^2}{[(k+\nu+1)(n+m+\nu+2)]^2}\Big).
\end{eqnarray}
\end{proposition}
{\bf Proof.} The proof is immediate by using (\ref{tote}) and (\ref{5dpp}). $\square$

\begin{proposition}
\label{prosp4o}
Consider $ |\phi_n^{(\nu,\beta)}\rangle$ and $ |\phi_n^{(m+1, \nu,\beta)}\rangle$ two  states in the Hilbert space $\mathcal{H}$ .
The  operators  $B_mB_m^\dag$ and  $B_m^\dag B_m$ mean-values are given by
\begin{eqnarray}
\label{4d8}
\langle B_mB_m^\dag  \rangle_{\phi_n^{(m+1,\nu,\beta)}}&=&\Big[(\pi\hbar L^{-1})^{m+1}\mathcal{M}(n,m;\nu,\beta)\Big]^2,\\
\label{poty}
\langle B_m^\dag B_m \rangle_{\phi_n^{(\nu,\beta)}}&=&\Big[(\pi\hbar L^{-1})^{m+1}\mathcal{M}(n-m-1,m;\nu,\beta)\Big]^2,
\end{eqnarray}
where  $\langle  A_{\nu,\beta} \rangle_{\phi_n^{(\nu,\beta)}}:=\int_0^L dx\;\overline{\phi_{n}^{(\nu,\beta)}(x)}A_{\nu,\beta}\phi_{n}^{(\nu,\beta)}(x)$. 
\end{proposition}
{\bf Proof.} It uses Proposition \ref{propae3}. $\square$

\begin{corollary}
\label{col2}
The operators ${_{m+1}\Lambda}_{n,\nu,\beta}$ and ${^{n+1}\Theta}_{m,\nu,\beta}$
satisfy the following identities:
\begin{eqnarray}
\label{deux}
\fl
\langle{_{m+1}\Lambda}_{n,\nu,\beta}\,{_{m+1}\Lambda}_{n,\nu,\beta}^\dag\rangle_{\phi_n^{(n+1,\nu,\beta)}}=\Big[(\hbar\pi L^{-1})^{n-m}\Big]^2\mathcal{N}(n,n;\nu,\beta)\mathcal{N}^{-1}(n,m;\nu,\beta),\\
\label{lave}
\fl
\langle{_{m+1}\Lambda}_{n,\nu,\beta}^\dag{_{m+1}\Lambda}_{n,\nu,\beta}\rangle_{\phi_n^{(m+1,\nu,\beta)}}=\Big[(\hbar\pi L^{-1})^{n-m}\mathcal{M}(m,n;\nu,\beta)\mathcal{M}^{-1}(n,m;\nu,\beta)\Big]^2,\\
\label{des}
\fl
\langle {^{n+1}\Theta}_{m,\nu,\beta} {^{n+1}\Theta}_{m,\nu,\beta}^\dag\rangle_{\phi_n^{(n+1,\nu,\beta)}}=\Big[(\hbar\pi L^{-1})^{m-n}\Big]^2\mathcal{N}^{-1}(n,n;\nu,\beta)\mathcal{N}(n,m;\nu,\beta),\\
\label{oux}
\fl
\langle  {^{n+1}\Theta}_{m,\nu,\beta}^\dag{^{n+1}\Theta}_{m,\nu,\beta}\rangle_{\phi_n^{(m+1,\nu,\beta)}}=\Big[(\hbar\pi L^{-1})^{m-n}\mathcal{M}(n,m;\nu,\beta)\mathcal{M}^{-1}(m,n;\nu,\beta)\Big]^2,
\end{eqnarray}
where $\mathcal{N}(n,m;\nu,\beta)$ is given by
\begin{eqnarray}
\fl
\mathcal{N}(n,m;\nu,\beta)
=\prod_{k=0}^{m}\frac{(2n-k+1)}{(2n+2\nu+k+3)^{-1}}
\Big(1+\frac{\beta^2}{[(k+\nu+1)(2n+\nu+2)]^2}\Big)
\end{eqnarray}
and $\mathcal{N}(n,n;\nu,\beta)=\mathcal{M}^2(n,n;\nu,\beta).$
\end{corollary}

Remark that the equations (\ref{des}) and (\ref{oux})  can be obtained by replacing $\mathcal{N}$ and  $n,m$ by  $\mathcal{N}^{-1}$ and  $m,n$ in  (\ref{deux}) and (\ref{lave}),   respectively. 

\section{Coherents states}
Let $|\zeta_z^{[m,\nu,\beta]}\rangle, z\in\mathbb{C}$  be the eigenstates of the operator $A_{m,\nu,\beta}$ associated to the eignevalue $z$. Then, 
\begin{eqnarray}
\label{toto}
|\zeta_z^{[m,\nu,\beta]}\rangle=\mathcal{R}
\exp \Big(\frac{zx}{\hbar}\Big)
\phi_0^{(m,\nu,\beta)}(x),\quad \forall \,x\,\in[0,L],
\end{eqnarray}
where $\mathcal{R}$ is the normalization constant.
In order to determine  $\mathcal{R}$, let us consider the set $\mathcal{K}=\Big\{(q,p)|q\in[0,L], p\in\mathbb{R}\Big\}$ which corresponds to the classical phase space of the P\"oschl-Teller problem. We re-express the operator ${\bf A}_{m,\nu,\beta}$ in terms of ${\bf Q}$ and ${\bf P}$
i.e  ${\bf A}_{m,\nu,\beta}=W_{m,\nu,\beta}({\bf Q})+i{\bf P}, $  where their actions on the function $\phi$ are given by ${\bf Q}:\phi(x)\rightarrow x\phi(x)$ and ${\bf P}:\phi(x)\rightarrow -i\hbar\phi'(x)$ on $\mathcal{D}_{A}$. Latter on, we change the variable $z$ as  $z=W_{m,\nu,\beta}(q)+ip$  \cite{berg, gazeau} i.e $|\zeta_{W_{m,\nu,\beta}(q)+ip}^{[m,\nu,\beta]}\rangle=|\eta_{q,p}^{[m,\nu,\beta]}\rangle.$ Then, the equation (\ref{toto}) becomes 
\begin{eqnarray}
\label{totoer}
|\eta_{q,p}^{[m,\nu,\beta]}\rangle=\mathcal{R}_m^{(\nu,\beta)}(q)
\exp \Big(\frac{(W_{m,\nu,\beta}(q)+ip)}{\hbar}x\Big)
\phi_0^{(m,\nu,\beta)}(x), \, \forall \,x\in[0,L],
\end{eqnarray}
where $\phi_0^{(m,\nu,\beta)}$ is given in (\ref{tote}).
The  normalization constant $\mathcal{R}_m^{(\nu,\beta)}(q)$    is given by 
\begin{eqnarray}
\label{prposttt}
\mathcal{R}_m^{(\nu,\beta)}(q)
=\exp\Big(-\frac{LW_{m,\nu,\beta}(q)}{2\hbar}\Big)\widetilde{\mathcal{O}}_m(q;L;\nu,\beta)
,
\end{eqnarray}
where $\widetilde{\mathcal{O}}_m(q;L;\nu,\beta)$ is provided by the expression
\begin{eqnarray}
\label{wide}
\fl
\widetilde{\mathcal{O}}_m^2(q;L;\nu,\beta)&=&
\sum_{k=0}^m   \frac{(-m,-m-2\nu-1)_k}{(-m-\nu-\frac{i\beta}{\nu+m+1})_kk!\Gamma(m+\nu+2-k+\frac{i\beta}{\nu+m+1})}\cr
&\times&\sum_{s=0}^m\frac{(-m,-m-2\nu-1,)_s\Gamma(2m+2\nu-s-k+3)}{ (-m-\nu+\frac{i\beta}{\nu+m+1})_s s! \Gamma(m+\nu+2-s-\frac{i\beta}{\nu+m+1})}\cr
&\times&\Bigg[\sum_{k=0}^m \frac{(-m,-2\nu-m-1)_k}{(-m-\nu-\frac{i\beta}{\nu+m+1})_k\Gamma(m+\nu+2-k+i(\nu+m+1)\cot\frac{\pi q}{L}) k!       }\cr
&\times&\sum_{s=0}^m\frac{(-m,-2\nu-m-1)_s\Gamma(2m+2\nu-s-k+3)}{(-m-\nu+\frac{i\beta}{\nu+m+1})_s\Gamma(m+\nu+2-s-i(\nu+m+1)\cot\frac{\pi q}{L})s!}\Bigg]^{-1}.
\end{eqnarray}
For computational details, see Appendix C. 

In the limit, when the parameter $m\rightarrow 0$, the  coherent states (\ref {totoer}), (\ref{prposttt}) are reduced to  ones obtained by  Bergeron et al  \cite{berg}.

The scalar product of two coherents states  $|\eta_{q,p}^{[m,\nu,\beta]}\rangle$ and $|\eta_{q',p'}^{[m,\nu',\beta']}\rangle$ satisfies
\begin{eqnarray}
\label{prposqtt}
\fl
\langle\eta_{q',p'}^{[m,\nu',\beta']}|\eta_{q,p}^{[m,\nu,\beta]}\rangle=Le^{\frac{L\alpha}{2\hbar}}\mathcal{R}_m^{(\nu',\beta')}(q')K_m^{(\nu',\beta')}\mathcal{R}_m^{(\nu,\beta)}(q)K_m^{(\nu,\beta)}\widetilde{\mathcal{T}}(m;\nu,\nu',\beta,\beta',\frac{ L\alpha}{2\pi\hbar}),
\end{eqnarray}
where $\alpha=W_{m,\nu,\beta}(q)+W_{m,\nu',\beta'}(q')+i(p-p')$ and
\begin{eqnarray}
\label{aaaa}
\fl
\widetilde{\mathcal{T}}(m;\nu,\nu',\beta,\beta',\frac{ L\alpha}{2\pi\hbar})&=&\Big(-m-\nu-i\frac{\beta}{m+\nu+1},-m-\nu'+i\frac{\beta'}{m+\nu'+1}\Big)_m\cr
&\times&\sum_{k=0}^m \frac{2^{-2m-\nu-\nu'-2}(-m,-m-\nu-\nu'-1)_k}{(-\nu-m-i\frac{\beta}{\nu+m+1})_k\Gamma(m+\frac{\nu+\nu'}{2}+2-k-i\frac{L\alpha}{2\pi\hbar}) k! m!      }\cr
&\times&\sum_{s=0}^m\frac{(-m,-m-\nu-\nu'-1)_s\Gamma(2m+\nu+\nu'+3-k-s)}{(-m-\nu'+i\frac{\beta'}{\nu'+m+1})_s\Gamma(m+\frac{\nu+\nu'}{2}+2-s+i\frac{L\alpha}{2\pi\hbar}) s!m!}.
\end{eqnarray}

\begin{proposition}
\label{labels}
The coherent states defined in (\ref{totoer}) 
\begin{enumerate}
\item are normalized
\begin{eqnarray}
\label{qqqq}
\langle\eta_{q,p}^{[m,\nu,\beta]}|\eta_{q,p}^{[m,\nu,\beta]}\rangle=1,
\end{eqnarray}
\item are not orthogonal  to each other,  i.e.
\begin{eqnarray}
\label{ppsqtt}
\langle\eta_{q',p'}^{[m,\nu',\beta']}|\eta_{q,p}^{[m,\nu,\beta]}\rangle\neq \delta(q-q')\delta(p-p'),
\end{eqnarray}
\item are continuous in $q,p$,
\item solve the identity, i.e.
\begin{eqnarray}
\label{uint}
\int_{\mathcal{K}}\frac{dq\,dp}{2\pi\hbar}|\eta_{q,p}^{[m,\nu,\beta]}\rangle\langle\eta_{q,p}^{[m,\nu,\beta]}|={\bf 1}.
\end{eqnarray}
\end{enumerate}
\end{proposition}
{\bf Proof.}

$\bullet$ Non orthogonality: From (\ref{prposqtt}) one can see that
\begin{eqnarray}
\label{prpttttt}
\langle\eta_{q',p'}^{[m,\nu',\beta']}|\eta_{q,p}^{[m,\nu,\beta]}\rangle\neq0,
\end{eqnarray}
which signifies  that the CS are not orthogonal.

$\bullet$ In the limit when the parameters $\nu'\rightarrow \nu,\beta'\rightarrow\beta,q'\rightarrow q$ and $p'\rightarrow p$, the quantity
\begin{eqnarray*}
Le^{\frac{L\alpha}{2\hbar}}\widetilde{\mathcal{T}}\Big(m;\nu,\nu',\beta,\beta',\frac{ L\alpha}{2\pi\hbar}\Big)\rightarrow\frac{1}{(\mathcal{R}_m^{(\nu,\beta)}(q)K_m^{(\nu,\beta)})^2} \mbox{ and }
\langle\eta_{q,p}^{[m,\nu,\beta]}|\eta_{q,p}^{[m,\nu,\beta]}\rangle=1,
\end{eqnarray*}
i.e the CS are normalized.

$\bullet$ Continuity in $q,p$
\begin{eqnarray}
\label{contu}
||(|\eta_{q',p'}^{[m,\nu,\beta]}\rangle-|\eta_{q,p}^{[m,\nu,\beta]}\rangle)||^2=2\Big(1-\mathcal{R}e\langle\eta_{q',p'}^{[m,\nu,\beta]}|\eta_{q,p}^{[m,\nu,\beta]}\rangle\Big).
\end{eqnarray}
So, $|||\eta_{q',p'}^{[m,\nu,\beta]}\rangle-|\eta_{q,p}^{[m,\nu,\beta]}\rangle||^2\rightarrow 0$
as $|q'-q|, |p'-p|\rightarrow 0$, since $\langle\eta_{q',p'}^{[m,\nu,\beta]}|\eta_{q,p}^{[m,\nu,\beta]}\rangle
\rightarrow 1$ as $|q'-q|, |p'-p|\rightarrow 0$.

$\bullet$ Resolution of the identity\\
Here we proceed as in \cite{berg} to show that
\begin{eqnarray}
\label{ui}
\fl
\int_\mathbb{R}\frac{4^{\nu+m}\Gamma(m+\nu+1-k+i(m+\nu+1)\cot\pi q)\Gamma(m+\nu+1-s-i(m+\nu+1)\cot\pi q)}{\pi^2\Gamma(2m+2\nu-k-s+2)}\cr
 \times\exp\Big((\nu+m+1)\cot\pi q (1-2x)\Big)dq\cr
=\frac{1}{\sin^{2\nu+2m+2}(\pi x)\Big(\frac{1+i\cot(\pi x)}{2}\Big)^k\Big(\frac{1-i\cot(\pi x)}{2}\Big)^s},\;\forall \,x \,]0, 1[,\;\forall \,\nu > -1.
\end{eqnarray}
 Let $\phi\in\mathcal{H}$ and $h_{q,p}$ a function in $ L^2(\mathbb{R},dx)$ defined by
\begin{eqnarray}
 h_{q,p}(x):=\left\{\begin{array}{l}\phi(x)\exp\Big(\frac{W_{m,\nu,\beta}(q)+ip}{\hbar}x\Big)
\phi_0^{(m,\nu,\beta)}(x),\quad \mbox{ if } x\in[0,L]\\
0\qquad\quad \qquad\quad\quad\quad\qquad\qquad\quad\qquad\mbox{ otherwise }.\end{array}\right.
\end{eqnarray}
One can see that the scalar product $\langle\eta_{q,p}^{[m,\nu,\beta]}|\phi\rangle$ given by
\begin{eqnarray}
\label{parpan}
\langle\eta_{q,p}^{[m,\nu,\beta]}|\phi\rangle=\mathcal{R}_m^{(\nu,\beta)}(q)\int_{0}^{L}dx\,e^{-i\frac{p}{\hbar}x}\phi(x)
\exp \Big(\frac{W_{m,\nu,\beta}(q)x}{\hbar}\Big)
\overline{\phi_0^{(m,\nu,\beta)}(x)}
\end{eqnarray}
is the Fourier transform of $h_{q,p},$ i.e. $\langle\eta_{q,p}^{[m,\nu,\beta]}|\phi\rangle=\mathcal{R}_m^{(\nu,\beta)}(q)\hat{h}_{q,p}(p/\hbar).$ Since  the
 function $h_{q,p}\in\,L^1(\mathbb{R},dx)\cap L^2(\mathbb{R},dx)$,  by using the Plancherel-Parseval  Theorem (PPT) we have
\begin{eqnarray}
\label{rrparpan}
\int_{\mathbb{R}}\frac{dp}{2\pi\hbar}|\langle\eta_{q,p}^{[m,\nu,\beta]}|\phi\rangle|^2=(\mathcal{R}_m^{(\nu,\beta)}(q))^2\int_{0}^L\frac{dx}{4\pi^2}|h_{q,p}(x)|^2.
\end{eqnarray}
The Fubini theorem yields
\begin{eqnarray}
\label{rrttparpan}
\fl
\int_{\mathcal{K}}\frac{dqdp}{2\pi\hbar}|\langle\eta_{q,p}^{[m,\nu,\beta]}|\phi\rangle|^2
=\int_0^Ldx \int_0^L\frac{dq}{4\pi^2}(\mathcal{R}_m^{(\nu,\beta)}(q))^2|\phi(x)|^2e^{\frac{2W_{m,\nu,\beta}(q)x}{\hbar}}\overline{\phi_0^{(m,\nu,\beta)}(x)}\phi_0^{(m,\nu,\beta)}(x)\cr
\end{eqnarray}
After using the inverse Fourier transform (see Appendix D), the above equation yields
\begin{eqnarray}
\int_{\mathcal{K}}\frac{dqdp}{2\pi\hbar}|\langle\eta_{q,p}^{[m,\nu,\beta]}|\phi\rangle|^2=\int_0^Ldx |\phi(x)|^2.
\end{eqnarray}
By using the polarization identity on the interval $[0, L]$, i.e $\int_{0}^Ldx|\psi(x)|^2=\int_{0}^Ldx\langle\psi|x\rangle\langle x|\psi\rangle$ we get the resolution of the identity. $\square$

\section{Conclusion}
In this paper, we have determined   a familly of normalized eigenfunctions of the hierarchic Hamiltonians of the P\"oschl-Teller Hamiltonian ${\bf H_{\nu,\beta}}$.  New operators with novel relevant properties and  their mean values  
 are determined. A new hierachic familly  of CS 
 is determined and discussed. In the limit, when $m\rightarrow 0$, the constructed CS well  reduce to the CS investigated  by Bergeron et al \cite{berg}.

\section*{Acknowledgements}
This work is partially supported by the Abdus Salam International
Centre for Theoretical Physics (ICTP, Trieste, Italy) through the
Office of External Activities (OEA) - \mbox{Prj-15}. The ICMPA
is also in partnership with
the Daniel Iagolnitzer Foundation (DIF), France.

\section*{Appendix A. The normalization constant of the eigenvector $|\phi_n^{(\nu,\beta)}\rangle$}
By using the property of the eigenstates, we have
\begin{eqnarray}
\label{eqnart}
\delta_{n,m}&=:&(\phi_{n}^{(\nu,\beta)},\phi_{m}^{(\nu,\beta)})\cr
&=&\overline{K_n^{(\nu,\beta)}}K_m^{(\nu,\beta)}\int_{0}^L dx \sin^{2\nu+n+m+2}\frac{\pi x}{L}\;
e^{-\frac{\beta \pi x}{L}\Big(\frac{1}{\nu+n+1}+\frac{1}{\nu+m+1}\Big)}\cr
&\times& \overline{P_n^{(a_n,\bar{a}_n)}\Big(i\cot \frac{\pi x}{L}\Big)}P_m^{(a_m,\bar{a}_m)}\Big(i\cot \frac{\pi x}{L}\Big)\cr
&=&\overline{K_n^{(\nu,\beta)}}K_m^{(\nu,\beta)}
\frac{(\bar{a}_n+1)_n(a_m+1)_n}{n!m!}\sum_{k=0}^m \frac{(-m,a_m+\bar{a}_m+m+1)_k}{(\bar{a}_m+1)_kk!}\cr
& \times& \sum_{s=0}^n \frac{(-n,a_n+\bar{a}_n+n+1)_s}{(a_n+1)_ss!}\times{\bf \mathcal{J}} 
   , 
\end{eqnarray}
where 
\begin{eqnarray*}
\fl
{\bf \mathcal{J}} =2^{-k-s}\int_{0}^L dx\Big[ \sin^{2\nu+m+n+2}\frac{\pi x}{L}
e^{-\frac{\beta \pi x}{L}\Big(\frac{1}{\nu+n+1}+\frac{1}{\nu+m+1}\Big)}\Big(1+i\cot \frac{\pi x}{L}\Big)^k
\Big(1-i\cot \frac{\pi x}{L}\Big)^s\Big]
\end{eqnarray*}
In  \cite{berg} it is shown that  
\begin{eqnarray*}
\int_{0}^1 dx \sin^{2\delta+2}(\pi x)e^{zx}=\frac{\Gamma(2\delta+3)e^{z/2}}{4^{\delta+1}\Gamma(\delta+2+i\frac{z}{2\pi})\Gamma(\delta+2-i\frac{z}{2\pi})}, \;\;\delta> -\frac{3}{2}.
\end{eqnarray*}
Therefore,
\begin{eqnarray*}
\fl
\frac{\delta_{n,m}}{\overline{K_n^{(\nu,\beta)}}K_m^{(\nu,\beta)}}
&=&L\frac{(-\nu-m-\frac{i\beta}{\nu+m+1})_m(-\nu-n+\frac{i\beta}{\nu+n+1})_n}{n!m!\exp\Big\{-\frac{\beta \pi }{2}\Big(\frac{1}{\nu+n+1}+\frac{1}{\nu+m+1}\Big)\Big\}2^{2\nu+n+m+2}}\cr
&\times&\sum_{k=0}^m   \frac{(-m,-2\nu-m-1)_k}{(-\nu-m-\frac{i\beta}{\nu+m+1})_kk!\Gamma(\frac{n+m}{2}+\nu+2-k+\frac{i\beta}{\nu+n+1})}\cr
&\times&\sum_{s=0}^n \Bigg\{\frac{(-n,-2\nu-n-1,)_s\Gamma(n+m+2\nu-s-k+3)}{ \Big(-\nu-n+i\frac{\beta }{2}\Big(\frac{1}{\nu+n+1}+\frac{1}{\nu+m+1}\Big)\Big)_s s! }\cr
&\times&\frac{1}{\Gamma\Big(\frac{n+m}{2}+\nu+2-s-i\frac{\beta }{2}\Big(\frac{1}{\nu+n+1}+\frac{1}{\nu+m+1}\Big)\Big)}\Bigg\}.
\end{eqnarray*}
The proof is achieved by taking $n=m.$

\section*{Appendix B.  Computation of $\phi_{n}^{(1,\nu,\beta)}$ }
From (\ref{eigen1}) we have
\begin{eqnarray}
\label{ddxdf}
\fl
\frac{d}{dx}\phi_{n+1}^{(\nu,\beta)}(x)
&=&\frac{\pi K_{n+1}^{\nu,\beta} }{L}\Bigg[\Big((\nu+n+2)\cos \frac{\pi x}{L}-\frac{\beta }{\nu+n+2}\sin \frac{\pi x}{L}\Big)\cr
&\times&P_{n+1}^{(a_{n+1},\bar{a}_{n+1})}\Big(i\cot \frac{\pi x}{L}\Big)+\frac{i(n+2\nu+2)}{2}
\sin^{-1}\frac{\pi x}{L}\\
&\times&P_{n}^{(a_{n+1}+1,\bar{a}_{n+1}+1)}\Big(i\cot \frac{\pi x}{L}\Big)\Bigg]\sin^{\nu+n+1}\frac{\pi x}{L}\exp \Big(-\frac{\beta \pi  x}{L(n+\nu+2)}\Big)\nonumber
\end{eqnarray}
and 
\begin{eqnarray*}
\fl
W_{\nu,\beta}\phi_{n+1}^{(\nu,\beta)}(x)
&=&\frac{\hbar \pi K_{n+1}^{\nu,\beta}}{L}\Bigg[\frac{\beta}{\nu+1}
\sin \frac{\pi x}{L}-(\nu+1)\cos \frac{\pi x}{L}\Bigg]\cr
&\times&\sin^{\nu+n+1}\frac{\pi x}{L}\exp \Big[\frac{-\beta \pi L^{-1}x}{L(\nu+n+2)}\Big]
P_{n+1}^{(a_{n+1},\bar{a}_{n+m+1})}
\Big(i\cot \frac{\pi x}{L}\Big).
\end{eqnarray*}
From the latter expression and
(\ref{ddxdf}),
we have
\begin{eqnarray*}
\fl
A_{\nu,\beta}\phi_{n+1}^{(\nu,\beta)}(x)&=&\frac{\pi \hbar K_{n+1}^{\nu,\beta}}{L}\Bigg[(n+1)\Big[\cos \frac{\pi x}{L}+\frac{\beta\sin \frac{\pi x}{L} }{(\nu+1)(\nu+n+2)}\Big]\cr
&\times&
P_{n+1}^{(a_{n+1},\bar{a}_{n+1})}\Big(i\cot \frac{\pi x}{L}\Big)
+\frac{i(n+2\nu+2)}{2}\sin^{-1}\frac{\pi x}{L}
\cr
&\times&P_{n}^{(a_{n+1}+1,\bar{a}_{n+1}+1)}
\Big(i\cot \frac{\pi x}{L}\Big)
\Bigg]\sin^{\nu+n+1}\frac{\pi x}{L}
\exp \Big(-\frac{\beta \pi  x}{L(n+\nu+2)}\Big).
\end{eqnarray*}
Let us determine $\cos \frac{\pi x}{L}+\frac{\beta }{(\nu+1)(\nu+n+2)}
\sin \frac{\pi x}{L}$:
\begin{eqnarray}
\label{init}
&&\cos \frac{\pi x}{L}+\frac{\beta }{(\nu+1)(\nu+n+2)}
\sin \frac{\pi x}{L}\cr
&=&\sqrt{1+\frac{\beta^2}{[(\nu+1)(\nu+n+2)]^2}}\Bigg[
\frac{1}{\sqrt{1+\frac{\beta^2}{[(\nu+1)(\nu+n+2)]^2}}}\cos \frac{\pi x}{L}\cr
&+&\frac{\beta}{(\nu+1)(\nu+n+2)\sqrt{1+\frac{\beta^2}{[(\nu+1)(\nu+n+2)]^2}}}\sin \frac{\pi x}{L}\Bigg]\cr
&=&\sqrt{1+\frac{\beta^2}{[(\nu+1)(\nu+n+2)]^2}}\Big(\cos \frac{\pi x}{L}\cos \alpha_{\nu,\beta}(n)+\sin \frac{\pi x}{L}\sin\alpha_{\nu,\beta}(n)\Big)\cr
&=&\sqrt{1+\frac{\beta^2}{[(\nu+1)(\nu+n+2)]^2}}\cos \Big(\frac{\pi x}{L}-\alpha_{\nu,\beta}(n)\Big)\cr
&=&\sqrt{\frac{\overline{\Delta_{n+1}^0 E_{0}^{(\nu,\beta)}}}{\varepsilon_0(2\nu+n+3)}}
\cos \Big(\frac{\pi x}{L}-\alpha_{\nu,\beta}(n)\Big).
\end{eqnarray}
Finally,
\begin{eqnarray*}
\fl
A_{\nu,\beta}\phi_{n+1}^{(\nu,\beta)}(x)
&=&K_{n+1}^{\nu,\beta}\Bigg[\sqrt{\frac{2M(n+1)^2\overline{\Delta_{n+1}^0 E_{0}^{(\nu,\beta)}}}{2\nu+n+3}}
\cos \Big(\frac{\pi x}{L}
-\alpha_{\nu,\beta}(n)\Big)\cr
&\times&P_{n+1}^{(a_{n+1},\bar{a}_{n+1})}\Big(i\cot \frac{\pi x}{L}\Big)
+\frac{i\pi\hbar(n+2\nu+2)}{2L\sin\frac{\pi x}{L}}P_{n}^{(a_{n+1}+1,\bar{a}_{n+1}+1)}\Big(i\cot \frac{\pi x}{L}\Big)\Bigg]\cr
&\times&\sin^{\nu+n+1}\frac{\pi x}{L}\exp \Big(-\frac{\beta \pi x}{L(\nu+n+2)}\Big),
\end{eqnarray*}
with 
\begin{eqnarray*}
\fl
\cos^2 \alpha_{\nu,\beta}(n)+\sin^2 \alpha_{\nu,\beta}(n)&=&\frac{1}{1+\frac{\beta^2}{[(\nu+1)(\nu+n+2)]^2}}+\frac{\beta^2}{\beta^2+[(\nu+1)(\nu+n+2)]^2}\cr
&=&1,
\end{eqnarray*}
and the constant 
$K_{n+1}^{\nu,\beta}$ defined in (\ref{prpos}).
\section*{Appendix C. Computation of the normalization constant of CS}
By using the definition
\begin{eqnarray}
\label{eqnarter}
&1=:(|\eta_{q,p}^{[m,\nu,\beta]}\rangle,|\eta_{q,p}^{[m,\nu,\beta]}\rangle)\cr
&=(\mathcal{R}_m^{(\nu,\beta)})^2\int_{0}^L dx \exp\Big(\frac{2W_{m,\nu,\beta}(q)x}{\hbar}\Big)\overline{\phi_0^{(m,\nu,\beta)}(x) }\phi_0^{(m,\nu,\beta)}(x) \cr
&=(\mathcal{R}_m^{(\nu,\beta)}K_m^{(\nu,\beta)})^2\int_{0}^L dx \sin^{2\nu+2m+2}\frac{\pi x}{L}
\exp \Big(\frac{2W_{m,\nu,0}(q)x}{\hbar}\Big)\cr
&\times\overline{P_m^{(a_m,\bar{a}_m)}\Big(i\cot \frac{\pi x}{L}\Big)}
 P_m^{(a_m,\bar{a}_m)}\Big(i\cot \frac{\pi x}{L}\Big)\cr
&=(\mathcal{R}_m^{(\nu,\beta)}K_m^{(\nu,\beta)})^2\frac{(\bar{a}_m+1)_m(a_m+1)_m}{(m!)^2}\sum_{k=0}^m \frac{(-m,a_m+\bar{a}_m+m+1)_k}{(\bar{a}_m+1)_kk!}\cr
& \times \sum_{s=0}^m \frac{(-m,a_m+\bar{a}_m+m+1)_s}{(a_m+1)_ss!}\times{\bf \mathcal{S}} 
   , 
\end{eqnarray}
with
\begin{eqnarray*}
{\bf \mathcal{S}} =2^{-k-s}\int_{0}^L dx\Bigg\{\sin^{2\nu+2m+2}\frac{\pi x}{L}
\exp \Big(-\frac{2\pi (\nu+m+1)x}{L}\cot\frac{\pi q}{L}\Big)\cr
\Big(1+i\cot \frac{\pi x}{L}\Big)^k
\Big(1-i\cot \frac{\pi x}{L}\Big)^s\Bigg\}\cr
=2^{-k-s}Le^{\frac{i(k-s)\pi}{2}}\int_{0}^1 dx \sin^{2\delta+2}(\pi x)e^{tx},
\end{eqnarray*}
In  \cite{berg} it is shown that  
\begin{eqnarray*}
\int_{0}^1 dx \sin^{2\delta+2}(\pi x)e^{tx}=\frac{\Gamma(2\delta+3)e^{t/2}}{4^{\delta+1}\Gamma(\delta+2+i\frac{t}{2\pi})\Gamma(\delta+2-i\frac{t}{2\pi})}, \;\;\delta> -\frac{3}{2}.
\end{eqnarray*}
Then the relation (\ref{eqnarter}) becomes
\begin{eqnarray*}
\fl
\frac{1}{(\mathcal{R}_m^{(\nu,\beta)})^2}
=&L(K_m^{(\nu,\beta)})^2(2^{\nu+m+1}m!)^{-2}e^{-\pi(\nu+m+1)\cot\frac{\pi q}{L}}\Big|\Big(-\nu-m+i\frac{\beta}{\nu+m+1}\Big)_m\Big|^2\cr
&\sum_{k=0}^m \frac{(-m,-2\nu-m-1)_k}{(-m-\nu-\frac{i\beta}{\nu+m+1})_k\Gamma(m+\nu+2-k+i(\nu+m+1)\cot\frac{\pi q}{L}) k!       }\cr
&\sum_{s=0}^m\frac{(-m,-2\nu-m-1)_s\Gamma(2m+2\nu-s-k+3)}{(-m-\nu+\frac{i\beta}{\nu+m+1})_s\Gamma(m+\nu+2-s-i(\nu+m+1)\cot\frac{\pi q}{L})s!},
\end{eqnarray*}
where $K_m^{(\nu,\beta)}$  is given in (\ref{prpos}).


\section*{Appendix D. Integral involved in  the resolution of the identity}
Here, in similar way as in  \cite{berg}, using the well-known Fourier transform (\cite{grad} p 520)
\begin{eqnarray}
\fl
\forall\,k\in\mathbb{R}, \forall\;\nu>-1, \quad \int_\mathbb{R}
\frac{e^{-itx}}{2\pi\cosh^{2\delta+2}(x)}dx=\frac{4^\delta\Gamma(\delta+1-i\frac{t}{2})\Gamma(\delta+1+i\frac{t}{2})}{\pi\Gamma(2\delta+2)}.
\end{eqnarray}
The inverse Fourier transform yields
\begin{eqnarray}
\fl
\forall\,x\in\mathbb{R}, \forall\;\delta>-1, \quad 
\int_\mathbb{R}\frac{4^\delta\Gamma(\delta+1-i\frac{t}{2})\Gamma(\delta+1+i\frac{t}{2})}{\pi\Gamma(2\delta+2)}e^{ikx}dx=\frac{1}{\cosh^{2\delta+2}(x)}.
\end{eqnarray}
The analytical extension is unique;  then, the above equality can be extended for
$x\in\mathbb{C}$ with $\mathcal{I}m(x)\in\,]-\pi/2,\pi/2[.$  By taking $u=ix,t\rightarrow \frac{t}{\pi}$ and $ u=\pi x-\frac{\pi}{2},\;\delta=m+\nu-\frac{k}{2}-\frac{s}{2},\;t=-2\pi(\nu+m+1)\cot\pi q-i\pi(k-s)$, we arrive at
\begin{eqnarray*}
\fl
\int_\mathbb{R}\frac{\Gamma(m+\nu+1-k+i(m+\nu+1)\cot\pi q)\Gamma(m+\nu+1-s-i(m+\nu+1)\cot\pi q)}{\pi^2\Gamma(2m+2\nu-k-s+2)}\cr
\times \exp\Big((\nu+m+1)\cot\pi q(1-2x)\Big)dq\cr
=\frac{4^{-\nu-m}}{\sin^{2\nu+2m+2}(\pi x)\Big(\frac{1+i\cot(\pi x)}{2}\Big)^k\Big(\frac{1-i\cot(\pi x)}{2}\Big)^s},\; m+n+\nu-\frac{s}{2}-\frac{k}{2} >-1.
\end{eqnarray*}

\end{document}